\documentclass[%
 reprint,
 amsmath,amssymb,
 aps,prd, superscriptaddress
]{revtex4-2}

\usepackage{graphicx}
\usepackage{dcolumn}
\usepackage{bm}
\usepackage{hyperref}
\usepackage[[usenames,dvipsnames]{xcolor}
\usepackage{amsmath}
\usepackage{amssymb}
\usepackage{bm}
\usepackage{orcidlink}
\usepackage[capitalize]{cleveref}
\usepackage{float}
\usepackage{tikz}
\usetikzlibrary{quotes,angles}
\usepackage{orcidlink}

\newcommand{\multilinecell}[2][c]{%
  \begin{tabular}[#1]{@{}c@{}}#2\end{tabular}}

\newcommand{\gw}{\mathrm{GW}}
\newcommand{\SNR}{\mathrm{SNR}}
\newcommand{\Msun}{M_\odot}

\graphicspath{ {./Plots/} }

\crefname{section}{Sec.}{Secs.}
\begin{document}

\preprint{APS/123-QED}

\title{Next-generation global gravitational-wave detector network: Impact of detector orientation on compact binary coalescence and stochastic gravitational-wave background searches}

\author{Michael Ebersold\, \orcidlink{0000-0003-4631-1771}}
\affiliation{Laboratoire d'Annecy de Physique des Particules, CNRS, 9 Chemin de Bellevue, 74941 Annecy, France}

\author{Tania Regimbau}
\affiliation{Laboratoire d'Annecy de Physique des Particules, CNRS, 9 Chemin de Bellevue, 74941 Annecy, France}

\author{Nelson Christensen\, \orcidlink{0000-0002-6870-4202}}
\affiliation{Universit\'e C$\hat{o}$te d’Azur, Observatoire de la C$\hat{o}$te d’Azur, CNRS, Artemis, 06304 Nice, France}


\date{\today}

\begin{abstract}

Next-generation gravitational-wave detectors like the Einstein Telescope and Cosmic Explorer, currently in their preparatory phase, have the potential to significantly improve our understanding of astrophysics, cosmology, and fundamental physics.
We examine how the arm orientations of the proposed detectors influence the sensitivity of a combined Einstein Telescope--Cosmic Explorer network with respect to the sensitivity to the stochastic gravitational-wave background and compact binary coalescences, where measuring both gravitational-wave polarizations is favorable. 
We present a method to optimize the arm orientations in the network for these two targets, and also demonstrate how to achieve a balanced configuration for both stochastic background and compact binary coalescence searches. For five specific network configurations, we explicitly compare the sensitivity to the stochastic background and binary neutron star mergers. For the latter, we conduct Bayesian parameter estimation on the extrinsic parameters of a reference binary neutron star system to assess sky localization and distance estimation capabilities. These are illustrated through efficiency curves showing the fraction of events meeting sky localization and distance uncertainty criteria as a function of redshift.
Our findings suggest that globally coordinating efforts towards the next-generation gravitational-wave detector network is advantageous.

\end{abstract}

\maketitle

\section{Introduction}

Since the first detection of gravitational waves (GWs) nine years ago~\cite{GW150914:2016}, the Laser Interferometer Gravitational-wave Observatory (LIGO)~\cite{AdvancedLIGO:2014} in the United States and the Virgo observatory in Europe~\cite{VIRGO:2014} have recorded $\sim \mathcal{O} (100)$ compact binary coalescences (CBCs) during their first three observing runs, O1 to O3~\cite{GWTC-3:2021}. The ongoing fourth (O4) and upcoming fifth (O5) observing runs are expected to significantly expand the catalog of GW events over the next few years~\cite{Prospects-LVK:2013}. 
To achieve sensitivity levels much higher than what is currently anticipated by upgrading existing detectors, it will be necessary to construct new facilities equipped with extended interferometer arms and advanced detector technologies. 
The Einstein Telescope (ET)~\cite{Hild:2009,Punturo:2010,Hild:2010} is the project pushed by the European GW community, while in the United States Cosmic Explorer (CE)~\cite{Reitze:2019,Evans:2021} is the project for a next-generation GW observatory. Both envision an improvement in sensitivity by 1 order of magnitude as well as a broader bandwidth toward low and high frequencies. 
The observations with next-generation detectors will facilitate a range of scientific pursuits, including electromagnetic follow-up observations, multimessenger astronomy, studies of compact binary populations, investigations into ultradense matter, cosmological research, and tests of general relativity~\cite{Maggiore:2019,Kalogera:2021,Evans:2021,Ronchini:2022}.

The baseline design of ET is a single underground observatory to reduce seismic noise. It has a triangular shape consisting of three nested detectors, each comprises two interferometers (``xylophone" configuration), one with high laser power optimized for high frequencies, the other tuned to low frequencies operating at cryogenic temperatures. However, recently a comparison study considering alternative designs and their science output has been conducted~\cite{Coba:2023}. In particular, different variations of a 2L configuration, which consist of two separated L-shaped interferometers while keeping the concepts of the ET design, were compared to the triangle. In a follow-up study, these investigations were extended to a 2L configuration containing one underground detector and one on the surface~\cite{Iacovelli:2024}.

The baseline concept for CE consists of two widely separated L-shaped observatories in the U.S., one with 40 km arm length, the other with 20 km arms. In Ref.~\cite{Gupta:2023} the scientific potential of different network configurations involving CE has been examined. Other studies that assess the scientific performance of a potential future detector network were conducted in Refs.~\cite{Borhanian:2022,Iacovelli:2022}.

Many of the comparison studies mentioned before~\cite{Borhanian:2022,Gupta:2023,Coba:2023} have in common that the geometry of the investigated networks is fixed in advance. Then, to assess the performance of these networks regarding CBCs, a catalog of CBCs is generated with population synthesis codes and properties like detection rate, range and accuracy in parameter estimation are reported. The uncertainty in the measurement of various parameters in the waveform is quantified with codes like \texttt{gwbench}~\cite{Borhanian:2020}, \texttt{GWFish}~\cite{Dupletsa:2022} or \texttt{gwfast}~\cite{GWFAST:2022}, which are based on the Fisher matrix formalism. While this formalism is computationally practical, it relies on several assumptions and has well-known limitations~\cite{Vallisneri:2007,Rodriguez:2013}. Nonetheless, there is a study that does rely on full Bayesian parameter estimation, although in the context of the advanced detector era~\cite{Emma:2024}. Furthermore, significantly faster parameter estimation codes making use of neural networks could make this approach more feasible in the future~\cite{Dax:2021,Dax:2022,Andres-Carcasona:2023,Wong:2023}.

The present work differs in many aspects from previous studies. First, we want to optimize the detector networks in terms of metrics that depend solely on the geometry and the sensitivities of the individual interferometers in the network, before looking at CBCs. Second, we do not want to choose a specific population model, but rather present the performance as efficiency curves for a reference system. Third, we employ Bayesian parameter estimation to evaluate the binaries' sky locations and distances and their uncertainties.
Moreover, we will focus on a full ET + CE network with CE in its baseline design and ET either in the triangle or the 2L configuration. However, the methods utilized in this paper could be extended to any other possible network. 
The locations of future interferometers will most likely be chosen due to geographical and political reasons, they will probably not yield much room for optimization, thus we fix them to the same as in Refs.~\cite{Coba:2023,Gupta:2023}. What there might be more freedom with is the arm orientation of the interferometers. How especially the relative arm orientation between the interferometers in the network impacts different scientific scenarios is the main focus of this work.

In this paper we will concentrate the discussion on two primary targets of next-generation GW detector networks, CBCs and stochastic GW backgrounds. We also study the observational potential for a global GW detection network, namely ET and CE together. The influence of the network's geometry on its sensitivity to stochastic backgrounds is determined by the overlap reduction function (ORF)~\cite{Christensen:1992,Flanagan:1993,Christensen:1997}. Conversely, for CBC searches, it is essential to identify an analogous function that will allow for the optimization of the detector network without the necessity of performing parameter estimation across numerous events.
Since CBC studies rely on accurate waveform reconstruction to improve source characterization, it is helpful if a detector network is able to reconstruct both polarizations. The function we will employ for this purpose is the network alignment factor~\cite{Klimenko:2005,Sutton:2009,Raffai:2013,Usman:2018}. It is a measure of how well a network can detect both GW polarizations, which can be computed from the geometry and the sensitivity of the interferometers alone and is a function of sky direction. 
Using the ORF and the network alignment factor as metrics, we will show how the arm orientations of the individual detectors can be chosen such that the network performs optimally. Since the two metrics prefer almost orthogonal orientations, we also demonstrate how they can be combined to achieve a good balance.

In order to show the impact of different network alignment factors and therefore different arm orientations on key CBC aspects like detection efficiency, localization and distance measurement, we consider a reference binary neutron star (BNS) system on which we estimate these quantities using Bayesian parameter inference, specifically \texttt{PyCBC inference}~\cite{Biwer:2018}. 
This has the benefit that our results are population model independent, in exchange, distinct properties of a population are not considered. 

The paper is structured as follows: in~\cref{sec:metrics} we introduce the two metrics for stochastic and CBC searches along which we will optimize the detector networks. The optimization as well as all of the properties of the investigated detector configurations are outlined in~\cref{sec:Networks}. In~\cref{sec:Stochastic} we have a closer look at the sensitivity of the networks regarding stochastic searches, while in~\cref{sec:CBC} we present the performance of the different networks regarding CBCs on a reference BNS system. This includes the standard detection efficiency as well as localization and distance measurements. Even though localization is more important in the case of BNS to find an electromagnetic counterpart, we also sketch how the study could be extended to binary black holes (BBHs). Finally we summarize and conclude this work in~\cref{sec:Conclusion}.

\section{Metrics to determine detector orientation} \label{sec:metrics}

Optimizing a network of GW detectors involves a delicate balance between sensitivity for stochastic background searches and the ability to localize and infer the properties of GW signals from CBCs. 
For optimal detection of the stochastic GW background (GWB), detector alignment within the network is crucial. In contrast, for CBC signals, the ability of the network to measure both gravitational-wave polarizations is preferred. For two L-shaped detectors, this means that their arms should be at an angle of $45^\circ$ with respect to each other, which in turn significantly reduces the sensitivity to the stochastic background. 
Another important aspect is the capability of a GW network to precisely localize an event, which in fact depends most on the number of detectors in a network and the distances between them, and only to a lesser amount on their arm orientation~\cite{Fairhurst:2010}. Since we fix the number and locations of the detectors and just want to optimize their arm orientation, the localization aspect is not directly put into the optimization process. Nevertheless, we will study the ability for sky localization in~\cref{sec:Localization}.
We first describe a metric to choose the optimal network for the stochastic background and then present a metric to determine how well both polarizations can be measured.

\subsection{Sensitivity of a GW network to the stochastic background: Power-law integrated sensitivity curves}
\label{sec:PIcurves}

The stochastic GWB is the result of the incoherent sum of numerous unresolved GW signals~\cite{Christensen:2018,Caprini:2018}. Its strength is usually given in terms of its energy density per logarithmic frequency interval with respect to the critical energy density $\rho_c$ required to close the universe,
\begin{align}
    \Omega_\gw (f) = \frac{1}{\rho_c} \frac{d\rho_\gw}{d \ln f} \,,
\end{align}
where
\begin{align}
    \rho_c = \frac{3c^2 H_0^2}{8\pi G} \,,
\end{align}
and $H_0$ denotes the Hubble expansion rate today. 

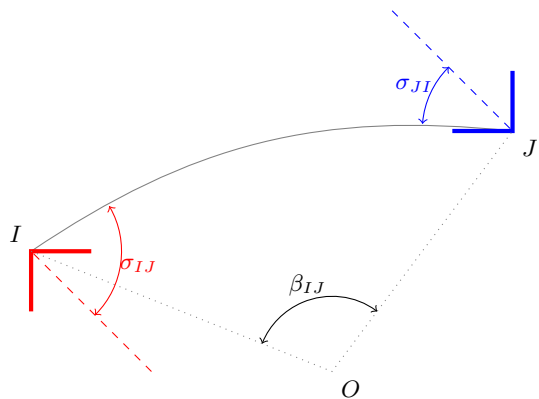
\begin{figure}[ht]
\begin{center}
    \begin{tikzpicture}[scale=0.8]
    \draw (8,2) coordinate (j) node[below right,black,thick] {$J$};
    \draw (0,0) coordinate (i) node[above left,black,thick] {$I$};
    \draw (5,-2) coordinate (o) node[below right, black,thick] {$O$};
    \draw (0.866,0.5) coordinate (ii);
    \draw (7.0,2.08) coordinate (jj);
    \draw[gray,thin, bend left=20] (i) edge (j);
    \draw[red,ultra thick] (0,-1) -- (i) -- (1,0);
    \draw[red,dashed] (2,-2) coordinate (ib) -- (i);
    \draw[blue,ultra thick] (7,2) -- (j) -- (8,3);
    \draw[blue,dashed] (6,4) coordinate (jb) -- (j);
    \draw[gray,dotted] (6,4) (i) -- (o) -- (j);
    \draw[red]
    pic["$\sigma_{IJ}$", draw=red, <->, angle eccentricity=1.2, angle radius=1.2cm]
    {angle=ib--i--ii};
    \draw[blue]
    pic["$\sigma_{JI}$", draw=blue, <->, angle eccentricity=1.2, angle radius=1.2cm] {angle=jb--j--jj};
    \draw[black]
    pic["$\beta_{IJ}$", draw=black, <->, angle eccentricity=1.2, angle radius=1.0cm] {angle=j--o--i};
\end{tikzpicture}
\end{center}
    \caption{Sketch of the relevant angles to compute the ORF for two detectors. $I$ and $J$ denote the two detector vertex locations on the surface connected by a great circle and $O$ indicates the center of Earth. $\beta_{IJ}$ is the angle between the two locations as seen from Earth's center. $\sigma_{IJ}$ is the angle between the bisector of detector $I$ and the connecting great circle towards detector $J$, measured in the tangential plane.}
    \label{fig:2detangles}
\end{figure}

The strategy to search for a stochastic background is to cross-correlate measurements of two or more detectors~\cite{Christensen:1992,Romano:2016}. This permits to eliminate the noise, which is assumed to be uncorrelated with the signal and between the detectors. However, in practice correlated environmental noise such as seismic or magnetic noise may remain~\cite{Thrane:2013-2,Thrane:2014,Janssens:2021}.
The expected signal-to-noise ratio (SNR) for a cross-correlation search for an unpolarized and isotropic stochastic background is given by~\cite{Allen:1997}
\begin{align} \label{eq:snr}
    \SNR_{IJ}=\sqrt{2 T}\left[\int_{f_{\min }}^{f_{\max }} d f \frac{\Gamma_{I J}^2(f) S_h^2(f)}{P_{n I}(f) P_{n J}(f)}\right]^{1 / 2} \,,
\end{align}
where $T$ is the total coincident observation time and $P_{nI} (f)$, $P_{nJ} (f )$ are the power spectral densities for the noise in detectors $I, J$. $S_h(f)$ refers to the strain power spectral density of the GWB, which is related to the more commonly used $\Omega_\gw(f)$ by
\begin{align} \label{eq:Sh}
    S_h(f)=\frac{3 H_0^2}{2 \pi^2} \frac{\Omega_\gw(f)}{f^3} \,,
\end{align}
and $\Gamma_{IJ}$ denotes the overlap reduction function (ORF).
It quantifies the reduction in sensitivity of the cross-correlation to a stochastic gravitational-wave background due to the response of the detectors and their separation and orientation relative to one another~\cite{Christensen:1992,Flanagan:1993,Christensen:1997}.
Given two detectors, the expectation value of the cross-correlation of the detector responses $\tilde{h}_I(f)$ and $\tilde{h}_J(f)$ is
\begin{align}
    \left\langle\tilde{h}_I(f) \tilde{h}_J^*\left(f^{\prime}\right)\right\rangle=\frac{1}{2} \delta\left(f-f^{\prime}\right) \Gamma_{I J}(f) S_h(f)\,,
\end{align}
and
\begin{align}
    \Gamma_{I J}(f) =\;& \frac{1}{8 \pi} \int d \hat \Omega \sum_A R_I^A(f, \hat{k}) R_J^{A *}(f, \hat{k}) \nonumber \\ 
    &\times e^{-i 2 \pi f \hat{k} \cdot\left(\vec{x}_I-\vec{x}_J\right) / c} \,,
\end{align}
where $\hat k$ is the traveling direction of the GW, $\vec x_I$ is the location of detector $I$ and $R_I^A (f,\hat{k})$ denotes the detector response to a sinusoidal plane GW of detector $I$ with polarization $A \in [+, \times]$.  The ORF is the transfer function between GW strain power and detector response cross-power. It is often convenient to define a normalized ORF $\gamma_{IJ}$ such that for two colocated and coaligned detectors, $\gamma_{IJ} (0) = 1$. Thus, for interferometers $I, J$ with opening angle $\theta_I$, $\theta_J$, between their arms,
\begin{align}
    \gamma_{IJ} (f ) = \frac{5}{\sin \theta_I \, \sin \theta_J} \Gamma_{IJ} (f ) \,.
\end{align}
The ORF for two detectors positioned on a spheroidal surface, such as the Earth, can be characterized using three angles which are given by the geometry of the detectors~\cite{Flanagan:1993}.
The first angle is $\beta_{IJ}$, which describes the angle between the two detectors $I, J$ and the center of the Earth. It can be calculated from the vectors pointing from the center of Earth to the vertex location of the detectors,
\begin{align}
    \beta_{IJ} = \arccos \left( \frac{\bm{r_I} \cdot \bm{r_J}}{|\bm{r_I}||\bm{r_J}|} \right)\,.
\end{align}
The other two angles are $\sigma_{IJ}$ and $\sigma_{JI}$ and denote the angle between the projection of the line joining the two detectors onto the plane of the individual detectors and the bisector of the two arms of the detectors. The directions (clockwise or anticlockwise) in which $\sigma_{IJ}$ and $\sigma_{JI}$ are chosen to be positive are unimportant, as long as the conventions coincide as $\beta \to 0$. A sketch with the detectors and the relevant angles is provided in~\cref{fig:2detangles}. 

It turns out that it is beneficial to express the ORF in terms of the combinations $\delta = \left(\sigma_{IJ}-\sigma_{JI}\right) / 2$ and $\Delta = \left(\sigma_{IJ}+\sigma_{JI}\right) / 2$, which can be interpreted as relative rotation and total rotation, respectively.
In addition, the ORF depends on the phase lag $\alpha = 2 \pi f |\bm{r_I} - \bm{r_J}|$ between the detectors, with $f$ being the frequency and $|\bm{r_I} - \bm{r_J}|$ the distance between the detectors. With these definitions we can write the ORF as
\begin{align} \label{eq:ORFabdD}
    \gamma_{IJ}(f) = \cos (4 \delta) \Theta_1(\alpha, \beta)+\cos (4 \Delta) \Theta_2(\alpha, \beta)\,,
\end{align}
where
\begin{subequations}
\begin{align} 
    \Theta_1(\alpha, \beta) =& \cos ^4(\beta / 2) g_1(\alpha) \,,\\
    \Theta_2(\alpha, \beta) =& \cos ^4(\beta / 2) g_2(\alpha) +g_3(\alpha)\nonumber\\
    &-\sin ^4(\beta / 2)\left[g_2(\alpha)+g_1(\alpha)\right]\,,
\end{align}
\end{subequations}
and the functions $g_j(\alpha)$ are all linear combinations of $\sin \alpha / \alpha^n$ and $\cos \alpha / \alpha^n$, given by
\begin{subequations}
\begin{align}
g_1(\alpha) &=\frac{5}{16} f(\alpha) \cdot(-9,-6,9,3,1)\,, \\
g_2(\alpha) &=\frac{5}{16} f(\alpha) \cdot(45,6,-45,9,3)\,, \\
g_3(\alpha) &=\frac{5}{4} f(\alpha) \cdot(15,-4,-15,9,-1)\,,
\end{align}
\end{subequations}
where
\begin{align}
f(\alpha) = \left(\alpha \cos \alpha, \alpha^3 \cos \alpha, \sin \alpha, \alpha^2 \sin \alpha, \alpha^4 \sin \alpha\right) / \alpha^5 \,.
\end{align}
From~\cref{eq:ORFabdD} we recognize that the absolute value of the ORF is periodic under the rotation of $\Delta$, $\delta$ and therefore $\sigma_{IJ}$, $\sigma_{JI}$ by $90^\circ$. As a result, the arm orientation of each individual detector can be shifted by $90^\circ$ without changing the ORF and thus the sensitivity to the stochastic background.

Since the ORF encapsulates the geometric factors that influence the cross-correlation signal between a detector pair, we quickly discuss the behavior of this function. It equals unity for coincident and coaligned detector and decreases below unity when the detectors are dislocated as there is a phase shift between the signals or if they no longer align with each other, making them sensitive to different polarizations. Therefore, in general, detectors should be close and coaligned to yield good sensitivity to the stochastic GWB, however, there are some more subtleties across the frequency range in the ORF~\cite{Flanagan:1993,Christensen:1997}. We also note that if two detectors are rotated with respect to one another by $45^\circ$, e.g. $\sigma_{IJ} = 45^\circ$ and $\sigma_{JI} = 0^\circ$, then $\delta = \Delta = 22.5^\circ$ and according to~\cref{eq:ORFabdD} we find $\gamma_{IJ} = 0$. Thus, such a detector pair does not contribute at all to the cross-correlation signal.
In~\cref{fig:ORF} we show graphically how the ORF looks for a few detector pairs that will be introduced in the next section. We refer to~\cref{tab:detectors} for the locations and orientations of these detectors. 

An elegant graphical representation of detector sensitivity curves for stochastic GWBs are power-law integrated (PI) sensitivity curves~\cite{Thrane:2013}. They take into account the increase in sensitivity that comes from integrating over frequency in addition to integrating over time. 
This method is valid for backgrounds that have a power-law spectrum in the analysis band of the form
\begin{align}
    \Omega_\gw (f) = \Omega_\beta \left(\frac{f}{f_\mathrm{ref}}\right)^{\beta} \,,
\end{align}
where $\beta$ is the spectral index and $f_\mathrm{ref}$ is an arbitrary reference frequency.

We quickly outline the construction of these curves following Ref.~\cite{Thrane:2013}. They are based on the expected SNR in a detector network, 
\begin{align} \label{eq:snrnet}
    \SNR = \sqrt{2 T}\left[\int_{f_{\min }}^{f_{\max }} d f \sum_{I=1}^M \sum_{J>I}^M \frac{\Gamma_{I J}^2(f) S_h^2(f)}{P_{n I}(f) P_{n J}(f)}\right]^{1 / 2} \,,
\end{align}
which is the generalization of~\cref{eq:snr} to more than two detectors. We need detector noise power spectral densities $P_{n I}$ of the detectors and the ORF between each two detectors, $\Gamma_{IJ}$. Then we can compute the effective strain power density 
\begin{align}
    S_{\mathrm{eff}}(f) \equiv\left[\sum_{I=1}^M \sum_{J>I}^M \frac{\Gamma_{I J}^2(f)}{P_{n I}(f) P_{n J}(f)}\right]^{-1 / 2}\,,
\end{align}
and convert it to energy density units using~\cref{eq:Sh},
\begin{align}
    \Omega_{\mathrm{eff}}(f) = \frac{2 \pi^2}{3 H_0^2} \, f^3\, S_{\mathrm{eff}}(f)\,.
\end{align}
We can compute the amplitude of $\Omega_\beta$ for a set of power-law indices $\beta$ and a reference frequency $f_{\mathrm{ref}}$ such that the integrated SNR has a fixed value $\SNR$ after an integration time $T$, explicitly
\begin{align}
    \Omega_\beta = \frac{\SNR}{\sqrt{2 T}}\left[\int_{f_{\min }}^{f_{\max }} d f \frac{\left(f / f_{\mathrm{ref}}\right)^{2 \beta}}{\Omega_{\mathrm{eff}}^2(f)}\right]^{-1 / 2} \,.
\end{align}
Usually one assumes $T=1$ yr and here we use a typical detection threshold of $\SNR=3$ corresponding to a false alarm probability of 10\% in Gaussian noise \cite{Romano_Cornish_2017}.
Note that the choice of $f_\mathrm{ref}$ is arbitrary and is not affecting the sensitivity curve.
Then for each pair of values for $\beta$ and $\Omega_\beta$ we plot $\Omega_\gw (f) = \Omega_\beta \left( f / f_\mathrm{ref} \right)^\beta $ versus $f$. The envelope of the $\Omega_\gw (f)$ power-law curves is the PI sensitivity curve for a correlation measurement using two or more detectors. Formally, the PI sensitivity curve is given by
\begin{align}
    \Omega_{\mathrm{PI}}(f) = \max _\beta\left[\Omega_\beta\left(\frac{f}{f_{\mathrm{ref}}}\right)^\beta\right]\, .
\end{align}
Any line (on a log-log plot) that is tangent to the PI sensitivity curve corresponds to a GWB power-law spectrum with $\SNR = 3$. This means that if the curve for a predicted background lies everywhere below the sensitivity curve, then $\SNR < 3$ for such a background. On the other hand, if the curve for a predicted power-law background with spectral index $\beta$ lies at any frequency above the sensitivity curve, then it will be observable with $\SNR = \Omega_\beta^\mathrm{pred} / \Omega_\beta > 3$ after a year of integration.

For any GW detector network, if the locations, orientations, and sensitivity curves of the individual detectors are fixed, the PI sensitivity curve $\Omega_\mathrm{PI}(f)$ can be computed. To compare networks, we will consider the minimum of the curve, which implies that we are comparing where the networks are most sensitive to a flat background. Depending on the science case, also different choices may be accurate. For instance, one could compare the tangential point of the curve to a certain power-law background or compare the value of $\Omega_\mathrm{PI}(f)$ at a fixed reference frequency.

\subsection{Metric for CBC searches: Network alignment factor}

We now introduce a metric that allows to optimize the geometry of a GW detector network such that the network is showing the best possible sensitivity to the second GW polarization across the sky.
In general, detector networks have varying sensitivity to the two polarizations. For a specific sky location, we can define the plus polarization as the linear combination to which the network is most sensitive. We can then determine the relative sensitivity to the cross polarization, known as the network alignment factor~\cite{Klimenko:2005}. In the following, we demonstrate how the network alignment factor is calculated for any interferometer network.

Consider a GW $h_+(t, \bm{x}), h_\times(t, \bm{x})$ from a sky direction $\hat \Omega$. The output of each detector $I \in  [1,\dots,D]  $ in a network of $D$ detectors is a linear combination of this signal and noise $n_I$:
\begin{align}
    d_I (t + \Delta t_I (\hat \Omega)) =&\; F_{I}^{+} (\hat \Omega) \, h_+ (t) + F_{I}^{\times} (\hat \Omega) h_\times(t) \nonumber\\ 
    &+ n_I (t + \Delta t_I(\hat \Omega)) \,.
\end{align}
Here $F^+ (\hat \Omega), F^\times ( \hat \Omega)$ denote the antenna response functions describing the sensitivity of the detector to the plus and
cross polarizations. Note that the choice of polarization basis is arbitrary. $\Delta t_I (\hat \Omega)$ refers to the time delay between the position of the detector and an arbitrary reference position, usually the center of Earth.
It is now convenient to define the noise-spectrum-weighted quantity
\begin{align}
F_{w,I}^{+,\times} (\hat \Omega, k) = \frac{F^{+,\times}_I}{\sqrt{\frac{N}{2} P_{nI} (k)}}\,,
\end{align}
where $k$ denotes the frequency bin and $N$ the number of data points. The $F_{w,I}^{+,\times} (\hat \Omega, k)$ are a function of sky position and frequency. Now for a network of detectors we can form the vectors
\begin{align}
    \bm{F^+} = 
    \begin{bmatrix} 
        F_{w,1}^{+} \\  F_{w,2}^{+}\\ \vdots \\F_{w,D}^{+} 
    \end{bmatrix} \,, \qquad \quad
        \bm{F^\times} = 
    \begin{bmatrix} 
        F_{w,1}^{\times} \\  F_{w,2}^{\times}\\ \vdots \\F_{w,D}^{\times} 
    \end{bmatrix} \,,
\end{align}
which contain all information on the sensitivity of the network as a function of frequency and sky position.

Since the choice of polarization is arbitrary and thus the direction of $\bm F^+$, $\bm F^\times$, it is possible to define it such that $\bm F^+$ points in the direction of maximal antenna response and $\bm F^\times$ in the direction the network has minimal antenna response. This choice of polarization definition is called the dominant polarization frame (DPF)~\cite{Klimenko:2005}. Often the polarization definition is chosen with reference to the source, in contrast, in the DPF it is defined with respect to the detector network at each frequency, which makes it very convenient to analyze different networks.

To construct the DPF explicitly, we first look at the antenna response in two frames separated by a polarization angle $\psi$~\cite{Anderson:2000}:
\begin{subequations}
\begin{align}
    \bm F^+ (\psi) &= \cos (2\psi) \bm F^+ (0) + \sin (2\psi) \bm F^\times (0) \,, \\
    \bm F^\times (\psi) &= -\sin (2\psi) \bm F^+ (0) + \cos (2\psi) \bm F^\times (0) \,.    
\end{align}
\end{subequations}
For any direction in the sky, one can always choose a polarization frame such that $\bm F^+ (\psi)$ and $\bm F^\times (\psi)$ are orthogonal and $ |\bm F^+ (\psi)| > |\bm F^\times (\psi)|$. The polarization angle which rotates $\bm F^+ (0) $ and $\bm F^\times (0)$ into the DPF can be computed by~\cite{Sutton:2009}
\begin{align}
    \psi_\mathrm{DPF} (\hat \Omega, k) =& \frac{1}{4} \mathrm{atan2} \big(2 \bm F^+ (0) \cdot \bm F^\times (0)), \nonumber\\
    &|\bm F^+ (0) |^2 - |\bm F^\times (0)|^2 \big) \,,
\end{align}
where $\mathrm{atan2(y,x)}$ is the arctangent function with range $(-\pi, \pi ]$. Let us denote the antenna response vectors in the DPF by $\bm f^+$, $\bm f^\times$. They have the properties $|\bm f^+|^2 \geq |\bm f^\times|^2$ and $\bm f^+ \cdot \bm f^\times = 0$. We can now define the network alignment factor as~\cite{Klimenko:2011}
\begin{align}
    \alpha = \frac{|\bm f^\times|}{|\bm f^+|} \quad \in [0, 1] \,. 
\end{align}
The alignment factor provides a measure of how sensitive a detector network is to the second polarization for each sky location. In~\cref{fig:alphasky} we will show the alignment factor in a Mollweide projection of the sky for two different detector networks.
Considering an isotropic distribution of GW sources, we can take the average over all sky locations as the metric to characterize a detector network. Moreover, because a good alignment factor is even more valuable if the network is sensitive in that particular direction, we weigh the average with the sensitivity $\sqrt{|\bm f^+|^2 + |\bm f^\times|^2}$ for each direction. 

In order to compare different GW networks, we can look at the weighted average alignment factor $\bar \alpha$, which depends on the location, orientation, and sensitivity curves of the individual detectors. Maximizing $\bar \alpha$ yields the network with the best sensitivity to the cross polarization.

\section{Detector Networks} \label{sec:Networks}

In this section we first define the detectors with their location and sensitivity curves. Then in a second step we compute their arm orientations such that the network optimizes the metrics defined in~\cref{sec:metrics}, for stochastic GWB and CBC searches. Additionally we search for an optimal trade-off.
Note that this concept of optimizing a GW detector network is by no means specific to our choices: any number of detectors, different locations, orientations and sensitivity curves can be examined. Here we consider a network of next-generation interferometers placed in four set locations and with fixed sensitivity curves, inspired by similar studies conducted in  Refs.~\cite{Coba:2023, Gupta:2023}, and mainly discuss the impact of their arm orientations. Moreover, we want to limit our investigations to a few network choices.

\subsection{Detector properties}

We start by describing the sensitivity curves and locations of the detectors in the networks we analyze throughout this paper. Although upgraded existing detectors like Virgo\_nEXT~\cite{Virgo_Next:2022} and LIGO $\mathrm{A}^{\sharp}$~\cite{ASharp:2022} could play a crucial role in a future GW detector network, here we look at the next-generation GW observatories, Einstein Telescope and Cosmic Explorer. Since the final geometry and locations are not yet fixed, the assumptions and designs investigated here should help in understanding how different geometries affect the capabilities of future GW detector networks. For all configurations we assume a Cosmic Explorer observatory in the United states and an Einstein Telescope observatory in Europe. CE is consisting of two L-shaped interferometers, one with 40 km arm length and one with 20 km arm length. For ET we consider two designs, first a triangular configuration with 10 km arm length, second two L-shaped interferometers with 15 km arm length each. 

\begin{figure}[ht]
    \centering
    \includegraphics[width=0.48\textwidth]{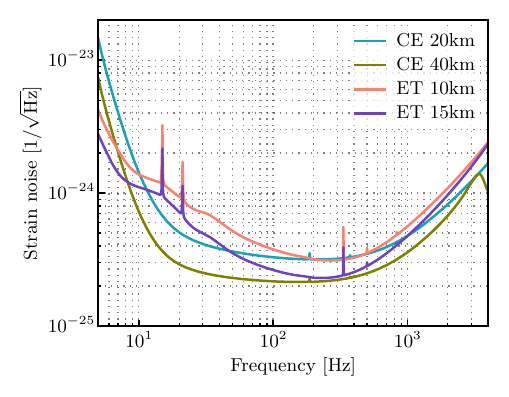}
    \caption{The strain noise curves of Einstein Telescope 10 and 15 km, as well as of Cosmic Explorer 20 and 40 km.}
    \label{fig:noisePSD}
\end{figure}

For the ET triangle, three detectors are nested into a triangular shape. Each of the detectors consists of a cryogenic low frequency and a high frequency interferometer (``xylophone design'')~\cite{Hild:2009}. For this study we assume a triangle of 10 km arm length and place it at the location of the Virgo observatory. This is the same design as was used, e.g., in Ref.~\cite{Regimbau:2012}. However, we use an updated sensitivity curve for a single 10 km long interferometer, which is provided in Ref.~\cite{ETSensitivityCurves}. Note that this noise curve is for a single interferometer and that the whole observatory with the three detectors is about a factor 3/2 more sensitive~\cite{Hild:2009}. The noise curve is displayed in~\cref{fig:noisePSD}.
The second ET design consists of two separated L-shaped interferometers. Here we assume 15 km long interferometers, one is placed at the candidate site in Sardinia, while the other is placed in the Meuse-Rhine three-border region across Belgium, Germany and the Netherlands. The exact locations are taken from Ref.~\cite{Coba:2023}. The elevation is chosen by estimating the altitude of the location and putting the interferometer approximately 200 m underground. The projected sensitivity of the 15 km interferometer is provided in Refs.~\cite{Coba:2023, ETSensitivityCurves} and shown in~\cref{fig:noisePSD}.

For Cosmic Explorer we adopt the sites and design described in Ref.~\cite{Gupta:2023}. A 40 km long interferometer is placed in the Pacific Ocean off the coast of Washington state and a 20 km long interferometer is placed in the Gulf of Mexico off the coast of Texas. The locations are intentionally chosen to be in the ocean to avoid impacting the ability to find a candidate site. The strain noise curves for the 40 km and the 20 km CE facilities are the same used in Refs.~\cite{Evans:2021, Gupta:2023}; they are provided in Ref.~\cite{CESensitivityCurves} and displayed in~\cref{fig:noisePSD}. In~\cref{tab:detectors} we summarize the locations of the considered detectors.

\subsection{Choice of arm orientation}

A key aspect of this work is to present the impact of the choice of orientation of the detector arms. Before we look at the network case with several interferometers, let us first consider only two L-shaped detectors and study their relative arm orientation. We will see that we can either maximize the sensitivity to the stochastic GWB by aligning their arms with respect to the great circle connecting the two sites, or we can put them at an angle of $45^\circ$ to maximize the sensitivity to the second polarization. 

Consider the ET L-shaped detectors at the sites in Sardinia and the Netherlands. From the discussion of the overlap reduction function we know that the best sensitivity towards the stochastic GWB is reached if the bisectors of the interferometers are pointing along the great circle towards the other site. In the tangential plane between local North and the direction of the great circle to the second site, this angle $\alpha_N$ is called azimuth and measured clockwise from local North. It can be calculated with the azimuth formula for a slightly squashed sphere, given the latitude, longitude of location 1 $(\theta_1, \lambda_1$) and location 2 $(\theta_2, \lambda_2$):
\begin{subequations} \label{eq:azimuth}
\begin{align}
    \tan \alpha_N = \frac{\sin (\lambda_2-\lambda_1)}{\left[\Lambda - \cos (\lambda_2 - \lambda_1) \right] \sin \theta_1 } \,,
\end{align}
where,
\begin{align}
    \Lambda &= (1-e^2) \frac{\tan \theta_2}{\tan \theta_1} + e^2\sqrt{\frac{1+(1-e^2)\tan^2 \theta_2}{1+(1-e^2)\tan^2 \theta_1}}\,, \\
    e^2 &= \frac{a^2 - b^2}{a^2} \,,
\end{align}
\end{subequations}
and $a$, $b$ denote the equatorial and polar radius of the Earth, respectively.
Here we use the cartographical definition of azimuth, i.e., the angle in the tangential plane where the detector lies is measured clockwise from North. This is the same definition used, e.g., in the \texttt{LALSuite} software~\cite{lalsuite}, but differs from other GW software like, e.g., \texttt{Bilby}~\cite{bilby-2019}, where the azimuth is measured counterclockwise from East.
Note that due to the curvature of Earth, the azimuth angle at the two ET candidate sites differs by $2.51^\circ$. This is expected, as the two detectors are not located in the same tangential plane. In other words, if one would use a simpler configuration and propose that at both locations the x arm of the detector points towards, e.g., East, the two detectors would not be aligned, but at an angle of $2.51^\circ$.

\begin{figure}[ht]
    \centering
    \includegraphics{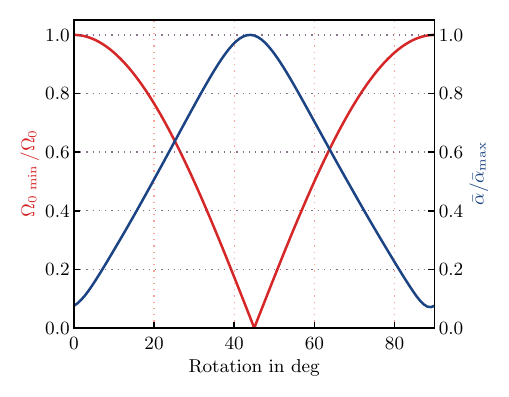}
    \caption{The network alignment factor and the minimum of the PI sensitivity curve are plotted as a function of the angle between the bisectors of the two L-shaped detectors at the two ET candidate sites. The red curve shows the stochastic metric relative to the minimal value, which is found if the two detectors are perfectly aligned. The blue curve represents the sky-averaged network alignment factor relative to the maximal value at 45 degrees.}
    \label{fig:alphaOmegaRot}
\end{figure}

For this configuration we can now compute the metrics described in~\cref{sec:metrics}, the minimum of the PI sensitivity curve, and the sky-averaged alignment factor. While fixing the detector in Sardinia and rotating the second detector around, we can plot how the metrics evolve for different orientations. We find the maximum of the sky-averaged alignment factor when the detectors are at 45 degrees, $\bar \alpha_{\max} = 0.737$, and the minimum of the PI sensitivity curve when the detectors are aligned, $\Omega_{\mathrm{PI,}\;{\min}} = 5.39 \times 10^{-13}$.  Figure~\ref{fig:alphaOmegaRot} provides a convenient illustration of how the two metrics behave opposed to each other. Note that, while the minimal value of $\bar \alpha$ is about 90 \% smaller than the maximum, two detectors at 45 degrees are completely blind to the stochastic GWB.

Let us now consider the global detector network consisting of an ET and a CE observatory. The locations and sensitivity curves of the interferometers stay fixed as described before, but we keep the orientation of the arms open. Since the sky-averaged network alignment factor and the minimum of the PI sensitivity curve are functions of the arm orientations, we can choose them such that the metric of interest is maximized/minimized or we can search for a trade-off.

\begin{table*}[ht]
\begin{tabular}{ |l|c|c|c|c|c|c|c| }
 \hline
 Detector name & Prefix & Arm length & Latitude & longitude & Elevation & x-arm azimuth & y-arm azimuth\\
 \hline
 Triangle-ET1 & E1 & 10 km & $43.63^\circ$ & $10.50^\circ$ & 52 m & $19.43^\circ$ & $319.43^\circ$ \\ 
 Triangle-ET2 & E2 & 10 km & $43.72^\circ$ & $10.55^\circ$ & 60 m & $259.43^\circ$ &  $199.43^\circ$\\ 
 Triangle-ET3 & E3 & 10 km & $43.70^\circ$ & $10.42^\circ$ & 60 m & $139.43^\circ$ & $79.43^\circ$ \\
 \hline
 ET-L-Sardinia & \multilinecell[c]{S1\\S2\\S3} & 15 km & $40.52^\circ$ & $9.417^\circ$ & 300 m & \multilinecell[c]{$100.60^\circ$ \\  $68.25^\circ$ \\ $54.90^\circ$ } & \multilinecell[c]{$10.60^\circ$ \\ $338.25^\circ$ \\ $324.90^\circ$} \\ 
 \hline
 ET-L-MeuseRhine & \multilinecell[c]{R1\\R2\\R3} & 15 km & $50.72^\circ$ & $5.921^\circ$ & 0 m & \multilinecell[c]{$188.14^\circ$ \\ $191.33^\circ$ \\ $170.99^\circ$} &  \multilinecell[c]{$98.14^\circ$ \\ $101.33^\circ$ \\ $80.99^\circ$ }\\ 
 \hline
 CE-Pacific & \multilinecell[c]{P1\\P2\\P3\\P4\\P5} & 40 km & $46.00^\circ$ & $-125.0^\circ$ & 0 m & \multilinecell[c]{$165.19^\circ$ \\ $181.90^\circ$ \\ $167.11^\circ$ \\ $161.82^\circ$ \\ $177.98^\circ$} & \multilinecell[c]{$75.19^\circ$ \\ $91.90^\circ$ \\ $77.11^\circ$ \\ $71.82^\circ$ \\ $87.98^\circ$}\\
 \hline
 CE-Atlantic & \multilinecell[c]{Y1\\Y2\\Y3\\Y4\\Y5} & 20 km & $29.00^\circ$ & $-94.0^\circ$ & 0 m & \multilinecell[c]{$4.82^\circ$ \\ $71.29^\circ$ \\ $81.52^\circ$ \\ $91.15^\circ$ \\ $70.06^\circ$} & \multilinecell[c]{$274.82^\circ$ \\ $341.29^\circ$ \\ $351.52^\circ$ \\ $1.15^\circ$ \\ $340.06^\circ$} \\ 
 \hline
\end{tabular}
\caption{Detector properties, positions, and arm orientations. The azimuth angle is measured clockwise from North.} \label{tab:detectors}
\end{table*}

\subsection{Analyzed networks}

In this study we want to look into the performance of a next-generation global detector network. Therefore we do not consider ET or CE alone, but rather examine the impact of the arm orientations when combining the detectors.
We consider five detector networks, each being a combination of CE with two distinct ET designs and different arm orientations. The first three combine CE with two 15 km L-shaped ET detectors, the last two with a 10 km triangle. The networks are listed in Table~\ref{tab:networks} together with the value of the two metrics. The orientation of the x and y arms of the detectors can be found in Table~\ref{tab:detectors}. Note that the detectors obey a 90 degree rotational symmetry, thus each individual detector in a network can be rotated by 90 degrees without impacting the performance of the network.

The first network \texttt{4L optimal $\Omega$} aims to optimize the search for the stochastic GWB, thus the arm orientations are chosen such that the PI sensitivity curve $\Omega_{\mathrm{PI,}\;{\min}}$ is minimized. Looking at the arm orientations, we see that the close detectors, i.e., the two ET detectors ($0.05^\circ$) and the two CE detectors ($0.25^\circ$) are almost exactly aligned to each other. Additionally, the ET detectors are approximately aligned to the more sensitive CE detector, $1.63^\circ$ and $4.25^\circ$.

The second network \texttt{4L optimal $\alpha$} intends to be best for CBC, thus the network alignment factor $\bar \alpha$ is maximized. We find $\bar \alpha = 0.83$ if the alignment between the ET detectors is chosen to be $35.6^\circ$ and between the CE detectors as $50.0^\circ$. As expected, it is closer to $45^\circ$, but the influence of other detectors is larger than for the optimal stochastic network. 
If instead we would fix the relative ET and CE orientations to $45^\circ$, depending on the relative angle between the two, we would find an alignment factor between 0.66 and 0.81, while drastically deteriorating the sensitivity towards a stochastic GWB. This shows that it is important to consider the full network.
A welcome side effect is that maximizing the alignment factor in a four detector network does not automatically mean that we get the worst sensitivity for the stochastic GWB as it was the case for a two detector network. 

\begin{figure}[ht]
    \centering
    \includegraphics[width=0.45\textwidth]{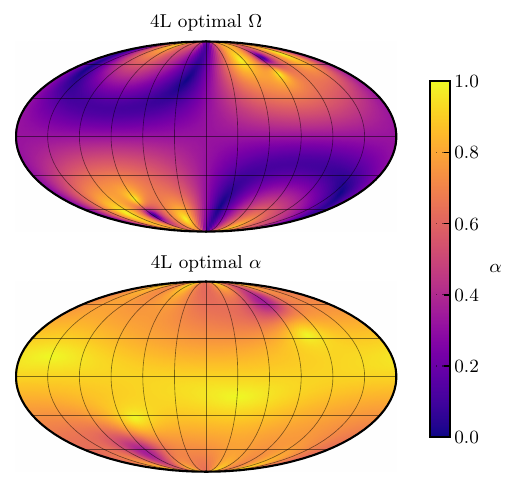}
    \caption{Mollweide projection of the sensitivity weighted network alignment factor for the detector configurations involving 4L-shaped interferometers which maximize sensitivity towards the stochastic background (\texttt{4L optimal $\Omega$}) and the second polarization (\texttt{4L optimal $\alpha$}).}
    \label{fig:alphasky}
\end{figure}

\begin{figure}[ht]
    \centering
    \includegraphics[width=0.48\textwidth]{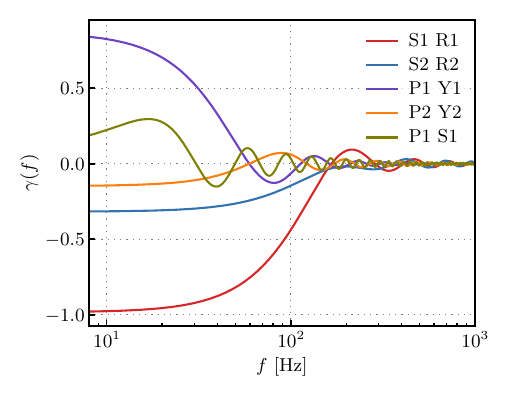}
    \caption{The overlap reduction function $\gamma (f)$ is shown for some detector pairs in the \texttt{4L optimal $\Omega$} and the \texttt{4L optimal $\alpha$} network. The negative values for $\gamma(f)$ as $f \to 0$ for some detector pairs are due to a rotation by $90^\circ$, which does not influence the absolute value of $\gamma$ and thus the sensitivity.}
    \label{fig:ORF}
\end{figure}

In order to visualize the difference between these two networks, we explicitly show the network alignment factor for all sky directions in~\cref{fig:alphasky} and in~\cref{fig:ORF} the ORF, which brings the network geometry into $\Omega_{\mathrm{PI,}\;{\min}}$.
While the second network has very good sensitivity to the second polarization almost over the whole sky, the first network shows larger areas where only one polarization is seen. This is of course expected as that is exactly what leads to a better measurement of the cross-correlation. Note also that not being sensitive to the second polarization does not mean not being sensitive to a GW signal. In some sky directions the measured SNR can indeed be higher.

In~\cref{fig:ORF} we illustrate the ORF for some detector pairs. Although the ET pairs S1, R1 and the CE pairs P1, Y1 are both almost perfectly aligned, the maximal value of the ORF for the CE pair is significantly lower due to the further distance between the two, $\sim 3000$ km against the $\sim 1000$ km between the ET pairs. This effect is even increased when looking at the P1, S1 pair. Moreover, the number of zeros per frequency range increases when the detectors are further apart due to the phase lag in the ORF [see~\cref{eq:ORFabdD}]. 
For the ET (S2, R2) and CE pairs (P2, Y2) in the \texttt{4L optimal $\alpha$} network, the maximal value of $\gamma(f)$ is much reduced due to the alignment which is closer to $45^\circ$. 

\begin{figure}[ht]
    \centering
    \includegraphics[width=0.48\textwidth]{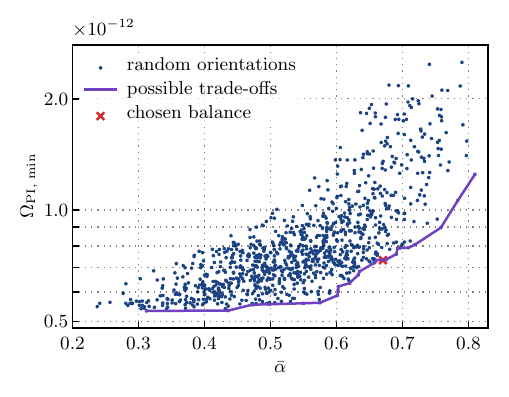}
    \caption{The minimum of the PI sensitivity curve and the network alignment factor is plotted for a 1000 random orientations of the four interferometers of ET and CE in the 4L configuration. On the envelope at the bottom right we find possible good trade-offs between the two metrics. The balance we choose for the \texttt{4L balanced $\alpha \Omega$} network is marked. }
    \label{fig:balance}
\end{figure}

For the third network, \texttt{4L balanced $\alpha\Omega$}, we search a trade-off between maximizing the alignment factor and minimizing the PI sensitivity curve. There is no unique way of defining an optimal trade-off. One should look for orientations such that any change that improves one metric necessarily deteriorates the other. This process is illustrated graphically in~\cref{fig:balance}. For 1000 random orientations of the four interferometers we plot the values of the two metrics on a plane. Since we want to maximize the alignment factor on the x axis and at the same time minimize the minimum of the PI sensitivity curve on the y axis, we find good trade-offs on the bottom right envelope. Then the optimal solution depends on the weights one wants to give to each metric. This choice is not unique and requires careful consideration, as the network alignment factor does not directly relate to an observable. Here, as a proof of principle, we choose the orientations such that $\Omega_{\mathrm{PI,}\;{\min}}$ is increased by about 38\% with respect to the \texttt{4L optimal $\Omega$} network and $\bar \alpha$ is decreased by about 19\% with respect to the \texttt{4L optimal $\alpha$} network.

In the two remaining networks, we combine the two CE detectors with the ET triangle. We keep the orientation of the triangle fixed and only rotate the CE detectors to minimize/maximize the metrics. Rotating the triangle has a minimal effect on the metrics due to its design with three nested detectors. Thus, for the fourth network, \texttt{$\Delta$ + 2L optimal $\Omega$}, we optimize the arm orientation of the CE detectors such that the PI sensitivity curve is minimized. Again we find that the two CE detectors should be aligned.
For the fifth network, \texttt{$\Delta$ + 2L optimal $\alpha$}, we rotate the CE detectors such that the alignment factor is optimized. The optimal angle between the two CE detectors is found to be $52.7^\circ$. 
Looking at the values of the metrics in~\cref{tab:networks}, we can see that the \texttt{$\Delta$ + 2L optimal $\Omega$} network, where the CE detectors are aligned, is significantly more sensitive to the stochastic background than the \texttt{$\Delta$ + 2L optimal $\alpha$} network, even though both contain the triangle. Because of the close interferometers, the triangle is quite sensitive to the stochastic background with $\Omega_{\mathrm{PI,}\;{\min}} = 2.3 \times 10^{-12}$. However, despite being further apart, we find for the aligned CE interferometers alone $\Omega_{\mathrm{PI,}\;{\min}} = 7.8 \times 10^{-13}$, due to the enhanced power spectral density (PSD) in most of the frequency range.
As for the 4L configuration, we could again think about a network that tries to balance the sensitivity to the stochastic background and the second polarization. Depending almost exclusively on the choice of the angle between the CE interferometers, such a network would perform in between the optimized networks \texttt{$\Delta$ + 2L optimal $\Omega$} and \texttt{$\Delta$ + 2L optimal $\alpha$}.

Note that while the minimum of the PI sensitivity curve is directly related to the sensitivity towards the stochastic background, the network alignment factor does not yield directly anything similar. This also means that while the $\Omega_{\mathrm{PI,}\;{\min}}$ can be directly compared for different networks, $\bar \alpha$ is merely a factor to compare different arm orientations in a network. Take, e.g., the ET triangle alone, it has an averaged network alignment factor of 0.84; however, adding two CE detectors does make the network much more sensitive although the alignment factor decreases to maximally 0.74.

\begin{table}[ht]
    \centering
    \begin{tabular}{|l|l|l|l|}
        \hline
        Network & Detectors & $\Omega_{\mathrm{PI},\;{\min}}$ & $\bar \alpha$ \\
        \hline
         4L optimal $\Omega$ & S1 R1 P1 Y1 & $5.3\times 10^{-13}$ & $0.37$ \\
         4L optimal $\alpha$ & S2 R2 P2 Y2 & $1.6\times 10^{-12}$ & $0.83$ \\
         4L balanced $\alpha\Omega$ & S3 R3 P3 Y3 & $7.3\times 10^{-13}$ & $0.67$ \\
        $\Delta$+2L optimal $\Omega$ & E1 E2 E3 P4 Y4 & $7.1\times 10^{-13}$ & $0.55$ \\
         $\Delta$+2L optimal $\alpha$ & E1 E2 E3 P5 Y5 & $1.5\times 10^{-12}$ & $0.74$ \\
    \hline
    \end{tabular}
    \caption{Considered detector networks: three configurations with four L-shaped interferometers, the first optimized for stochastic background searches (\texttt{4L optimal $\Omega$}), the second for CBC searches (\texttt{4L optimal $\alpha$}), and the third is a trade-off (\texttt{4L balanced $\alpha\Omega$}). Two configurations with ET as a triangle and the 2L of CE optimized once for stochastic searches (\texttt{$\Delta$+2L optimal  $\Omega$}) and once for CBC searches (\texttt{$\Delta$+2L optimal $\alpha$}). The table shows the individual detectors (see~\cref{tab:detectors}) in the network and respectively the minimum of the PI sensitivity curve and the sky-averaged network alignment factor.}
    \label{tab:networks}
\end{table}

\section{Searches for the stochastic GW background} \label{sec:Stochastic}

The stochastic GWB is formed by the incoherent superposition of signals emitted by different GW sources. One traditionally distinguishes between the astrophysical GWB (AGWB) and a cosmological GWB (CGWB). While the AGWB is due to the superposition of unresolved GWs from different astrophysical sources~\cite{Regimbau:2011}, the CGWB arises from processes in the early universe such as inflation, cosmological phase transitions or cosmic strings~\cite{Christensen:2018,Caprini:2018}. Searches for the GWB typically assume the background to be Gaussian, isotropic, stationary, and unpolarized, and look for excess correlated power between detector pairs~\cite{Allen:1997,Romano:2016}.
We first look at the sensitivity of our detector networks to an isotropic background, then we discuss their angular sensitivity and finish with prospects regarding the detection of the AGWB and the CGWB.

\subsection{Sensitivity to isotropic GW backgrounds}

\begin{figure}[ht]
    \centering
    \includegraphics[width=0.48\textwidth]{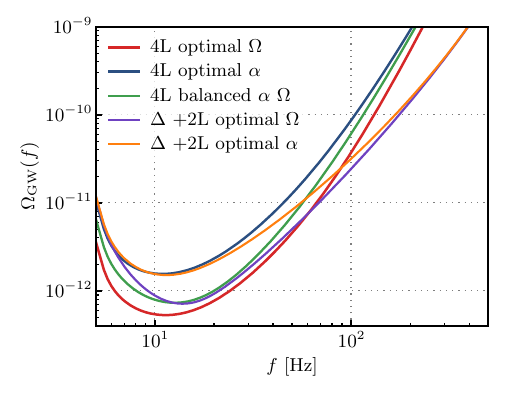}
    \caption{Power-law integrated sensitivity curves for future detector network's correlation search for power-law GWBs. The curves are interpreted as follows: A power-law background that is tangential to a sensitivity curve can be measured in one year of coincident observational data with a signal-to-noise ratio $\SNR = 3$.}
    \label{fig:PISensCurves}
\end{figure}

The sensitivity of a detector network to an isotropic GWB can be directly inferred from the PI sensitivity curves described in~\cref{sec:PIcurves}. They show the sensitivity of a network to a power-law background as a function of the geometry of the network, the noise PSDs of the individual detectors and the observation time. In~\cref{fig:PISensCurves}
we show the sensitivity curves for our five networks assuming 1 yr of observation time and a threshold $\SNR=3$.
Looking at the networks involving four L-shaped detectors, we observe that the minimum of the \texttt{4L optimal $\alpha$} network is about a factor of 3 larger than the minimum of the \texttt{4L optimal $\Omega$} network, which in turn means that in order to reach the same measurement sensitivity towards a flat background, one would need a 9 times longer observation time, i.e., nine years. With the \texttt{4L balanced $\alpha \Omega$} network, the same sensitivity is reached after roughly two years. For the two networks involving the triangle, the difference in observation time would be roughly 4.5 years.
We can also recognize that the sensitivity at higher frequencies is better for the networks involving the triangle. This is due to the fact that the overlap reduction function remains constant at high frequencies for colocated detectors.

\subsection{Angular sensitivity to the GW background}

The GWB is expected to show anisotropies due to the nature of spacetime along the line of sight, and certainly for the astrophysical contribution as a result of the local distribution of matter and the finiteness of the number of sources~\cite{Cusin:2017,Jenkins:2018,PhysRevLett.122.111101}. 
We study the angular sensitivity towards a stochastic GWB for the five network configurations following the formalism in Ref.~\cite{Alonso:2020}. 
Assuming that the dependence of the fractional energy density of a stochastic GWB on the frequency and the direction can be factorized, its statistical properties can be written in terms of a frequency-dependent angular power spectrum as
\begin{align} \label{eq:Cl}
    \langle \Omega_\gw (\hat n,f) \Omega_\gw (\hat n^\prime,f) \rangle = \sum_\ell C_\ell (f) P_\ell(\hat n \cdot \hat n^\prime) \,,
\end{align}
where $P_\ell (\hat n \cdot \hat n^\prime)$ are the Legendre polynomials, which are functions of the angle between the two directions $\hat n$ and $\hat n^\prime$.
The ability to measure a specific multipole component of the background intensity is quantified by $N_\ell$, the noise per multipole. It defines the SNR for an angular power spectrum as
\begin{align}
    \SNR^2 = \sum_\ell (2\ell + 1) \frac{C_\ell}{N_\ell} \,,
\end{align}
where the $C_\ell$ are the coefficients characterizing the angular distribution of power in the stochastic GWB across different scales as defined in~\cref{eq:Cl}.

\begin{figure}[ht]
    \centering
    \includegraphics[width=0.46\textwidth]{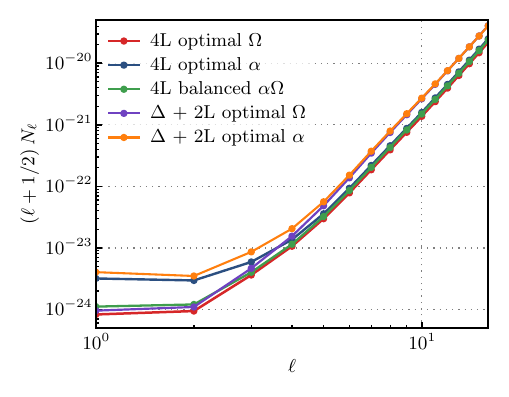}
    \caption{Noise per angular multipole at a reference frequency of $f_\mathrm{ref}=63$ Hz for the five detector networks.}
    \label{fig:Nell}
\end{figure}

Assuming that the noise is stationary in each detector, the $N_\ell$ are given by 
\begin{align}
    N_\ell^{-1} = \frac{T}{2} \sum_{ABCD} \int df \, G_\ell^{AB,CD}(f)\,,
\end{align}
where $T$ is the total observation time. $G_\ell^{AB,CD}(f)$ is a function of the angular multipole of the antenna patterns and the frequency dependence of the expected background (see Ref.~\cite{Alonso:2020}). In~\cref{fig:Nell} we show the noise per multipole $N_\ell$ at a reference frequency of $f_\mathrm{ref} = 63$ Hz as calculated with the \texttt{schNell} package~\cite{Alonso:2020,schNell:2024}. We can see that the networks optimized for stochastic searches are more sensitive at low multipoles, whereas for higher multipoles, $\ell > 5$, any 4L configuration is slightly superior than the networks with the ET triangle. This is due to the fact that having more detectors at different locations provides better angular resolution.
With the more favorable detector designs, we could start probing the anisotropy in the astrophysical background, which is expected at a level $(\ell +1/2) C_\ell \sim 10^{-25}$ at $f_\mathrm{ref} = 63$ Hz~\cite{Cusin:2018,Cusin:2019,Cusin:2019-2}. Moreover, the presence of Doppler anisotropies induced by the motion of the detectors with respect to the rest frame of the stochastic GWB source could be tested~\cite{Cusin:2022}.

\section{CBC Searches} \label{sec:CBC}

In this section we study the performance of CBC searches at the example of BNS signals with next-generation GW detector networks. Since we want to do that independently from specific astrophysical population model, we define a reference BNS system for which the detection efficiency, localization capability, and distance estimation of the different detector networks is presented. 
Thus, the main product is efficiency curves as a function of redshift, $\varepsilon(z)$.
They can then be multiplied with observation time $T$ and a merger rate $R(z)$ to get, e.g., the number of detected events. 
\begin{equation}
N_\mathrm{det}= T \int dz \, \varepsilon(z) R(z) 
\end{equation}
Note that we are not doing a full parameter estimation, thus we do not have results on, e.g., how well the chirp mass is measured. However, we expect it to depend little on the arm orientation and much more on the measured SNR. In contrast, we expect a more distinct difference in the distance and inclination measurements due to the different sensitivity to the second polarization.

\subsection{Detection efficiency}

We start by considering a fiducial population of equal mass BNS systems with each component being $1.4 \Msun$ in the source frame, which is distributed isotropically in orientation and sky location. The spins are assumed to be zero and the tidal deformability is modeled according to the APR4 equation of state~\cite{Akmal:1998}, thus for a neutron star of $1.4 \Msun$ we find $\Lambda_{1.4} = 257$. 
The BNS waveform is modeled using the phenomenological model \texttt{IMRPhenomXAS\_NRTidalv2}~\cite{Pratten:2020a,Dietrich:2019,Colleoni:2023} which models the inspiral and tapers the amplitude to zero above the estimated merger frequency. It has the leading $\ell=2$, $m=2$ mode available and accounts for tidal effects. A potential postmerger signal is not considered here.

To quickly determine whether a GW signal from a CBC is possibly detectable, we compute its SNR $\rho$. In a single detector A, the SNR is given by
\begin{align}
    \rho_A^2 = 4 \int^{f_h}_{f_l} df \, \frac{|\tilde{h}_A|^2}{P_{n A}} \,,
\end{align}
where $\tilde{h_A}$ is the GW strain in the frequency domain at detector $A$, $P_{n A}$ is the one-sided noise power spectral density (PSD) of detector $A$, and $f_l$ and $f_h$ denote lower and upper frequency cut of the integration.
For a network of detectors, we sum up the squares of the individual SNRs:
\begin{align}
    \rho = \sqrt{\sum_i\, \rho_i^2} \,.
\end{align}
Here we choose for the lower frequency 15 Hz, such that the signal has a length of approximately 340 s and therefore using a stationary antenna pattern function is a reasonable choice. Since the next-generation detectors are sensitive down to lower frequencies, we miss out on roughly 9 \% (12 \%) of the SNR if instead we would change the lower frequency to 10 Hz (5 Hz).

\begin{figure}[ht]
    \centering
    \includegraphics[width=0.46\textwidth]{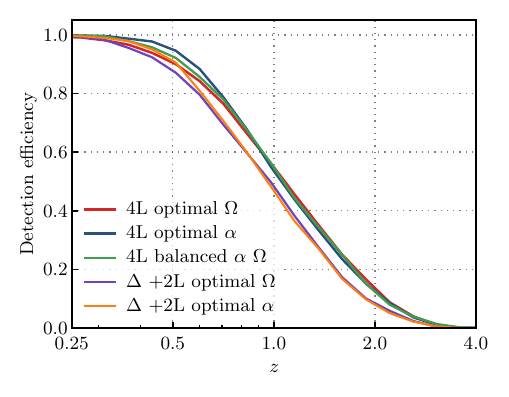}
    \caption{The network detection efficiency for the reference BNS system as a function of redshift is displayed for the five GW detector networks. It shows the fraction of BNS events detected with an SNR $\rho \geq 12$ at each redshift.}
    \label{fig:deteffBNS}
\end{figure}

The detection efficiency of a detector network is the fraction of all events at a redshift $z$ that are observed with an SNR greater than a certain threshold SNR. We set the network SNR threshold to $\rho = 12$ and additionally require that an individual detector has to observe the signal with a minimum SNR of $\rho = 5$ to contribute to the network SNR. The resulting efficiency curves for BNS systems are displayed in~\cref{fig:deteffBNS}. Up to a redshift of $\sim 0.3$ (1.6 Gpc), all networks detect every BNS coalescence. Then, the efficiency curves start to drop, first for the networks optimized for the stochastic GWB. Generally, the networks involving the triangular ET detector are a little less sensitive but still detect half the BNS systems at redshift 1. We can also recognize that at higher redshifts, $z > 0.8$, the detection efficiency is very little impacted by the orientation of the arms. Instead, the main differences arise where the detection efficiency starts to drop from 1. For the networks involving four L-shaped interferometers, we find that the one optimized for the alignment factor performs a little better than the balanced one, which itself performs a little better than the network optimized for stochastic searches. For example at $z=0.6$ (3.6 Gpc), \texttt{4L optimal $\alpha$} detects 88\% of all BNS coalescences, \texttt{4L balanced $\alpha\Omega$} 86\% and \texttt{4L optimal $\Omega$} 84\%.  

\begin{figure}[ht]
    \centering
    \includegraphics[width=0.46\textwidth]{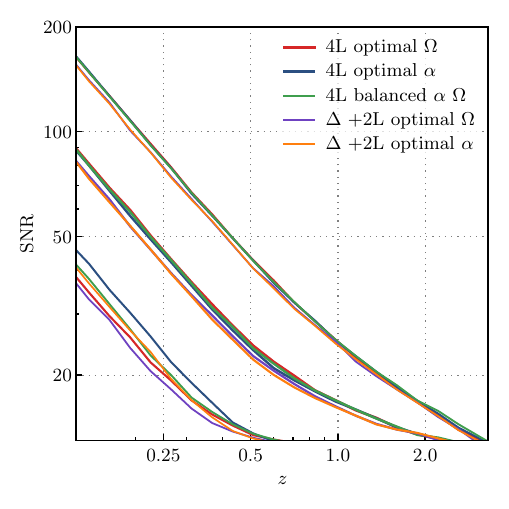}
    \caption{The SNR distribution of the detected reference BNS events is shown as a function of redshift for the five GW detector networks. The main line shows the median SNR at each redshift and the upper and lower lines include 90\% of the detected events.}
    \label{fig:snrBNS}
\end{figure}

In~\cref{fig:snrBNS} we indicate the SNR distribution of the detected events as a function of redshift for the five networks. The median SNR does not depend much on the arm orientation, it is very similar for the 4L networks and the networks involving the triangle, which are somewhat less sensitive. The main difference is in the lower 90\% credible interval, the networks with a better alignment factor $\bar \alpha$ show less events detected with a relatively small SNR.

\subsection{Sky localization} \label{sec:Localization}

A good sky localization of the source plays a crucial role in enabling multimessenger astronomy. The ability to accurately pinpoint the origin of GW signals allows for an efficient follow-up with electromagnetic (EM) telescopes, which allows them to capture EM transients that may accompany the binary merger.
Wide-field telescopes, such as the Vera Rubin Observatory~\cite{LSST:2008}, are expected to detect kilonovae associated with BNS coalescences below a redshift of about 0.4~\cite{Gupta:2023-2}. Beyond this redshift, only the emission in the high-energy band through beamed gamma-ray bursts (GRBs) is expected to be observed~\cite{GW170817MMA:2017}. 
Another important aspect of next-generation GW detectors is the ability to detect BNS inspirals minutes before the merger and send premerger alerts with a good estimate of the sky localization of the source~\cite{Nitz:2021,Chatterjee:2022,Hu:2023}. In this study we do not focus on the premerger localization, but instead consider the full signal. However, we expect that the relative difference between the networks in their localization capabilities would be similar for premerger signals.

There are several methods to determine the sky localization of GW signals. Conventional parameter estimation methods like \texttt{LALInference}~\cite{Veitch:2014}, \texttt{BILBY}~\cite{Ashton:2018} or \texttt{PyCBC Inference}~\cite{Biwer:2018} implement Bayesian inference for CBCs~\cite{Christensen:2022}. They use Markov chain Monte Carlo and nested sampling techniques to sample the entire parameter space. By marginalizing over nuisance parameters, one can obtain the source directions. These codes usually take $\sim$hours to run. A much faster method is \texttt{Bayestar}~\cite{Singer:2015}, a rapid Bayesian position reconstruction algorithm, which is used in following up GW detections and provides a sky map within several seconds.
For our comparison study, the full parameter estimation is computationally too expensive. We also find that \texttt{Bayestar} is not self-consistent especially for high SNR signals as it tends to underestimate the uncertainty, meaning that the true location is less than $90 \%$ of times within the $90 \%$ confidence area. Moreover, it is also dependent on the analyzed network, thus the comparisons would become flawed.

As an alternative, we use the single template model from \texttt{PyCBC Inference}~\cite{Biwer:2018, pycbc:2024}. It is only for extrinsic parameter estimation, which increases the speed by avoiding waveform generations while computing likelihoods. A single reference signal with fixed intrinsic parameters is generated. The likelihoods are precalculated up to constant factors which vary with the extrinsic parameters. This method relies on the fact that the posteriors of intrinsic and extrinsic parameters largely decouple for nonprecessing CBCs~\cite{OShaughnessy:2013,Singer:2015,Islam:2022}.  
For the purpose of comparing the sky localization and distance measurement of different detector networks, we assume the intrinsic parameters of a signal are known, which is analogous to the SNR time series used for \texttt{Bayestar}~\cite{Singer:2015}. 
However, neglecting the weak correlations between intrinsic and extrinsic parameters could lead to a slight underestimation of the uncertainties in the extrinsic parameters~\cite{Duverne:2023}.
The extrinsic parameters that are sampled are the luminosity distance~$d_L$, inclination angle~$\iota$, right ascension~$\alpha$, declination~$\delta$, polarization~$\psi$, coalescence phase~$\phi_c$, and coalescence time~$t_c$. 
Sampling of the parameter space is done using the \texttt{dynesty}~\cite{Dynesty:2020} sampler, a nested sampling algorithm. We assume standard isotropic priors for the sky location and the orientation angles of the source.
Although the next-generation detectors will be sensitive to GW down to 5 Hz or even below, here we start the inference analysis at 15 Hz.
We generate the signal using the \texttt{IMRPhenomXAS\_NRTidalv2} waveform model as above and place it into simulated detector noise which is assumed to be stationary Gaussian noise drawn from the respective noise curve.

\begin{figure}[ht]
    \centering
    \includegraphics[width=0.46\textwidth]{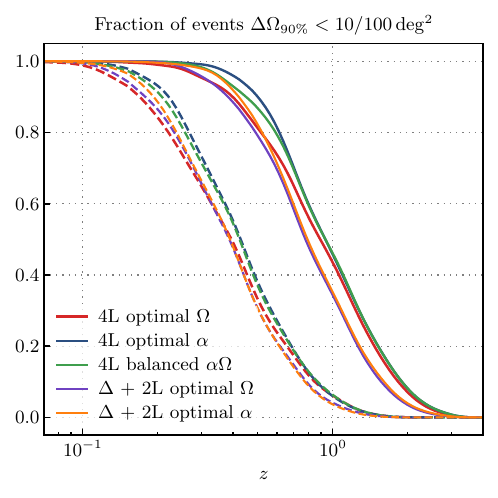}
    \caption{The fraction of BNS events with a sky localization $\Delta \,\Omega_{90\%} < 10 \mathrm{deg}^2$ (dashed lines) and $\Delta \Omega_{90\%} < 100 \,\mathrm{deg}^2$ (solid lines).}
    \label{fig:loceffBNS}
\end{figure}

We simulate $\mathcal{O}(10^5)$ GW signals from BNSs.
The signals are generated isotropically distributed in comoving volume, though we generate them in batches with different maximal distances, in order to get enough signals also at close distances.
If an event has a network SNR  $\rho \geq 12$, we consider it as detected and run the parameter estimation which yields posterior distributions for the seven extrinsic parameters. We can then convert the posterior distributions of right ascension and declination into a sky localization area; typically one uses the area in which the true event is at 90\% certainty, $\Delta \Omega_{90 \%}$.
In~\cref{fig:loceffBNS} we plot the fraction of BNS events localized within 10 and 100 $\mathrm{deg^2}$ as a function of redshift. 
For instance, it is evident that the best performing network, \texttt{4L optimal $\alpha$}, detects roughly 54\% of all BNS events at $z=1$ according to~\cref{fig:deteffBNS} and localizes 47\% within 100 $\mathrm{deg}^2$ according to~\cref{fig:loceffBNS}, or 87\% of the detected events. Similarly, about 6.3\% are localized within 10 $\mathrm{deg}^2$, which are $\sim 12\%$ of the detected events. Comparing these numbers to the balanced network,\texttt{4L balanced $\alpha \Omega$}, we find that they are not much different: at $z=1$, 55\% of all BNS events are observed with an SNR $\rho \geq 12$, 47\% are localized within 100 $\mathrm{deg}^2$, and 6.7\% within 10 $\mathrm{deg}^2$.
Even the network including the triangle optimized for stochastic searches, \texttt{$\Delta$+2L optimal $\Omega$}, is not much worse: at redshift $z=1$ about 48\% of all events are detected, roughly 36\% are localized within 100 $\mathrm{deg}^2$, and 4.4\% within 10 $\mathrm{deg}^2$. 

\begin{figure}[ht]
    \centering
    \includegraphics[width=0.46\textwidth]{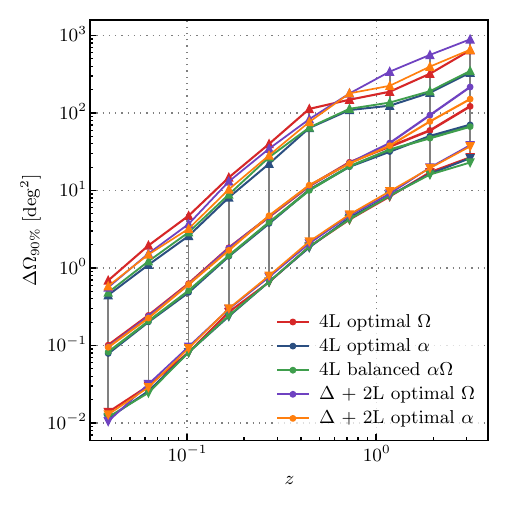}
    \caption{This plot shows the median, as well as the fifth and 95th percentile of the 90\% sky localization area $\Delta \Omega_{90\%}$ for some redshift bins.}
    \label{fig:locareas}
\end{figure}

To also get a feeling about the size and distribution of the 90\% sky localization areas, see~\cref{fig:locareas}. It provides a glimpse at the distribution of 90\% confidence areas at certain redshifts. While the median is very similar over most of the range for all networks, notably at the upper end of confidence areas at each redshift we can recognize differences. Particularly, the networks optimized for stochastic searches show a wider distribution, meaning more events are not so well localized. 
In general, we can see that the localization capability of any of the considered networks is very good. For instance, all networks can pinpoint more than 50\% of the events below $z=0.1$ to less than 1 $\mathrm{deg}^2$. 

\subsection{Distance measurements}

An important property of CBCs is that one can measure the luminosity distance to the source from their GW signal~\cite{Schutz:1986}, thus they are called ``standard sirens". However, the GW signal does not provide the redshift, therefore to constrain the cosmological parameters one needs to independently measure the redshift, e.g., through the association of the source to a host galaxy. This is possible through a direct EM counterpart or through statistical methods based on spatial correlations with galaxies and large-scale structures or on features in the mass distribution of the sources~\cite{LIGOCosmo:2021,Holz:2005,MacLeod:2007,Nissanke:2009,Messenger:2011,DelPozzo:2015}.
The better the measurement of the luminosity distance from the GW signal, the better all these methods work. An intrinsic problem is that there is a strong degeneracy with the determination of the orbital inclination angle with respect to the line of sight. This degeneracy usually leads to large uncertainties in the distance estimation, though, if the GW network is sensitive to both polarizations, the error can be significantly reduced~\cite{Usman:2018,Chassande-Mottin:2019}. Other methods to reduce the degeneracy from the GW signal alone involve higher-order modes or precession of the orbital plane, which are effects that are most likely suppressed in BNS systems due to the rather symmetric masses and small, aligned spins~\cite{London:2017,Vitale:2018,Giacomazzo:2010,Tauris:2017}. In the case of an EM counterpart, the inclination angle could be constrained independently from a GRB jet~\cite{Ziaeepour:2019,Pian:2020} or an observed afterglow light curve~\cite{Hotokezaka:2018}.

\begin{figure}[ht]
    \centering
    \includegraphics[width=0.46\textwidth]{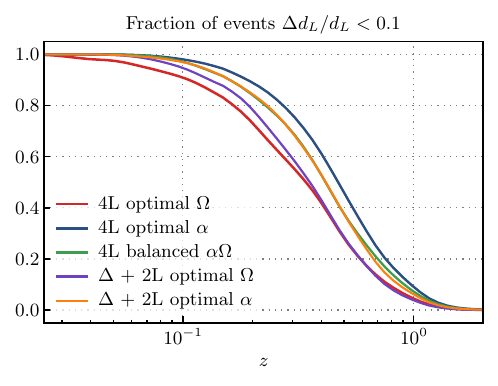}
    \caption{The fraction of events with a fractional uncertainty in distance smaller than 10\%.}
    \label{fig:disteffBNS}
\end{figure}

The parameter estimation runs using the single template model from \texttt{PyCBC Inference} described in~\cref{sec:Localization} also provide distance and inclination posteriors. To yield a measure of how well the networks can measure the distance, we present an efficiency curve as a function of redshift showing the fraction of events detected with a fractional error $\Delta d_L / d_L < 0.1$ in~\cref{fig:disteffBNS}. For $\Delta d_L$ we convert the 90\% credible interval on the distance posterior into approximately $1 \sigma$ error: $\Delta d_L = (\Delta_{95} d_L - \Delta_{5} d_L)/2/1.645$, where $\Delta_{5} d_L$ and $\Delta_{95} d_L$ represent the lower and upper limit of the 90\% credible interval, respectively. This is to simplify comparisons with other studies and works well if the distance posterior is sufficiently Gaussian, which is mostly the case. However, in certain cases, the posterior has more support at larger distances due to the prior distribution used.

The efficiency curves in~\cref{fig:disteffBNS} clearly show that GW networks that are more sensitive to the second polarization are performing better at measuring the distance. For instance, the \texttt{4L optimal $\alpha$} network detects more than 60\% of the events with a relative distance uncertainty below 10\% within a redshift of $z=0.4$. Comparing to the other 4L configurations at this redshift, the \texttt{4L optimal $\Omega$} network performs roughly 29\% worse, whereas the balanced network \texttt{4L balanced $\alpha \Omega$} is only about 14\% worse.
The difference between the configurations with the triangle is less pronounced, the network \texttt{$\Delta$+2L optimal $\Omega$} performs about 18\% worse than the network \texttt{$\Delta$+2L optimal $\alpha$} at a redshift of 0.4.
Another interesting feature in the plot is that when comparing the networks optimized for stochastic searches, the network containing the ET triangle is doing better below a redshift of 0.4, where all events are detected. This is due to the unique design of the triangle, which allows for a better measurement of the second polarization, and thus, has a better chance of breaking the distance-inclination degeneracy leading to a better distance measurement.

\begin{figure}[ht]
    \centering
    \includegraphics[width=0.46\textwidth]{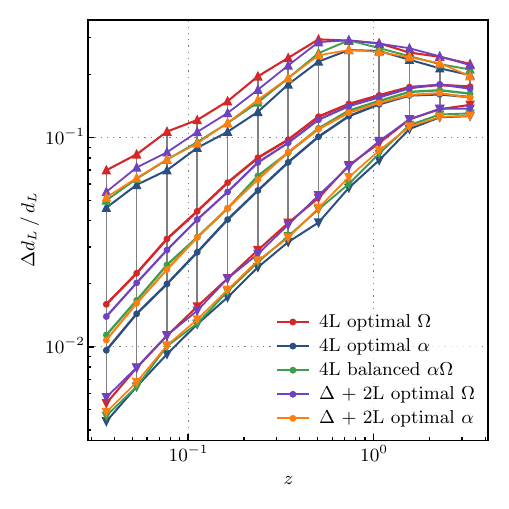}
    \caption{The plot shows the median, as well as the fifth and the 95th percentile of the relative distance uncertainty $\Delta d_L/d_L$ in some redshift bins for the five networks.}
    \label{fig:disterrs}
\end{figure}

The size of the relative distance uncertainty for the detected BNS events is displayed in~\cref{fig:disterrs} as a function of redshift for the five GW detector networks. As expected, the closer the event, the better the distance can be measured. This trend saturates at a relative distance measurement error of about 20\% for far away events, $z>1$, due to many signals not being detected anymore.
The best performing network, \texttt{4L optimal $\alpha$}, can measure the luminosity distance for more than 50\% of very close events $z<0.04$ to a subpercent precision.

In~\cref{tab:CBCresults} the results for detection efficiency, sky localization and distance estimation of the five networks are compactly recapitulated. It shows the fraction of BNS systems that are observed obeying the criteria for localization and distance estimation at two redshift values, $z = 0.4$ and $z=1.0$. 

\begin{table*}[ht]
\begin{tabular}{ |l|c|c|c|c|c|c| }
 \hline
  & \multicolumn{3}{c|}{$z = 0.4$} & \multicolumn{3}{c|}{$z = 1.0$} \\
 \hline
 Detector network & $\rho > 12$ & $\Delta \Omega_{90\%} < 10 \; (100) \,\mathrm{deg}^2$ & $\Delta d_L / d_L < 0.1 $ & $\rho > 12$ & $\Delta \Omega_{90\%} < 10 \; (100) \,\mathrm{deg}^2$ & $\Delta d_L / d_L < 0.1 $ \\
 \hline
 4L optimal $\Omega$ & 0.95 & 0.50 \,(0.90) & 0.44 & 0.55 & 0.064 \,(0.45) & 0.043 \\
 4L optimal $\alpha$ & 0.98 & 0.56 \,(0.96) & 0.63 & 0.54 & 0.063 \,(0.47) & 0.086 \\
 4L balanced $\alpha\Omega$ & 0.97 & 0.55 \,(0.93) & 0.54 & 0.55 & 0.067 \,(0.47) & 0.070 \\
 $\Delta$+2L optimal $\Omega$ & 0.94 & 0.49 \,(0.89) & 0.44 & 0.48 & 0.044 \,(0.35) & 0.035 \\
 $\Delta$+2L optimal $\alpha$ & 0.97 & 0.49 \,(0.92) & 0.54 & 0.47 & 0.039 \,(0.36) & 0.060 \\
 \hline
\end{tabular}
\caption{Overview of the CBC results for the five detector networks at two distinct redshift values. The table shows the fraction of events satisfying the indicated criteria for detection, localization and fractional distance uncertainty.} \label{tab:CBCresults}
\end{table*}

\subsection{Extension to binary black hole mergers}

We have evaluated the capabilities of next-generation detector networks for detecting, localizing, and measuring distances in CBC searches, using the standard reference BNS system of $1.4+1.4\Msun$, commonly employed to estimate detector sensitivity. Extending this analysis to BBHs and possibly neutron star-black holes presents no conceptual or technical challenge, however, some subtleties must be considered. Nonetheless, we do not expect the conclusions of our comparative study to differ significantly across different source types.


For BBHs we expect a more diverse mass distribution than for BNS, both in terms of total mass and mass ratio. Additionally, while spins for BNS are generally expected to be small and aligned with the orbital angular momentum~\cite{Tauris:2017}, this is not generically true for BBHs. These factors significantly enrich the morphology of the emitted GW signal and make a single reference system insufficient. Therefore, examining several reference models within this parameter space may be useful. 
Furthermore, while localization and distance estimation remain important for BBHs, accurately measuring their spins might be even more valuable, as it can yield insights into their formation history~\cite{Mandel:2009,Rodriguez:2016,Farr:2017,Gerosa:2018}. However, to evaluate how well detector networks can reconstruct the spins of BBHs, full parameter estimation would be necessary. 
In addition, due to the possible asymmetric masses and precession, higher modes in the GW waveform become important and can contribute considerably to the parameter estimation. In particular, the distance measurement gets more precise due to breaking the distance-inclination degeneracy~\cite{London:2017,Mills:2020,Liu:2024}. Also, additional features like GW memory can enhance parameter estimation and have been shown to be important for high SNR signals in next-generation detectors~\cite{Xu:2024}.

These complications lead to the fact that parameter estimation using the single template model is inappropriate and a full inference of all parameters is needed. However, this would increase the computational cost of a similar study enormously. Nonetheless, one could circumvent this by looking, e.g., at a few reference systems only in particular sky directions and apply likelihood reweighting techniques~\cite{Payne:2019}.

\section{Conclusion} \label{sec:Conclusion}

In this work we compared the performance of different configurations of a combined ET--CE detector network, focusing on the impact of detector arm orientation. While fixing the locations and power spectral densities of the interferometers, we presented a strategy to choose their arm orientations by optimizing a metric that can be computed purely by the geometry of the network. We have introduced two metrics, the minimum of the PI sensitivity curve and the sky averaged network alignment factor. The first optimizes a network for stochastic GWB searches whereas the second is a measure of how effective the network is in measuring both GW polarizations, from which particularly CBC searches profit.
Then we utilized these metrics to compute the arm orientations in five networks, two including ET triangle + CE ($\Delta$ + 2L) and three ET 2L + CE configurations (4L), such that they are optimized towards one metric or represent a good balance.
For these detector networks, we compared the PI sensitivity curves in detail and additionally discussed their angular sensitivity. Concerning CBCs, we have assessed the capabilities of these networks regarding detection, sky localization and distance measurement using Bayesian parameter estimation on a BNS reference system. This allows for a population independent comparison between the networks.

Although we particularly investigated the full ET--CE network, it is pivotal to note that the method of optimizing arm orientations in a detector network, utilizing metrics that incorporate the geometry and sensitivities of the individual detectors, works for any potential network, and can also include existing facilities.

Besides optimizing the arm orientations utilizing a single metric, we have introduced a balanced 4L network configuration, \texttt{4L balanced $\alpha \Omega$}. We have shown that it is possible to find a configuration that can significantly benefit one science target without substantially weakening the other. This network detects the same number of events and has nearly the same localization capabilities as the CBC optimized network, \texttt{4L optimal $\alpha$}, while it performs roughly 10\% weaker in distance measurement. In contrast, its sensitivity to the stochastic GWB is doubled. 

An important takeaway of this paper is that the next-generation GW detector network should be planned globally, potentially also including existing observatories. Coordinating the arm orientations, within site specific limitations, can substantially enhance the performance to different science targets. For instance, we have demonstrated that in a 4L configuration, which is designed to be sensitive to both polarizations, there is a better option than aligning the two ET and the two CE detectors at $45^\circ$. This alternative, our \texttt{4L optimal $\alpha$} network, is far more sensitive to the stochastic GWB than the configuration where both observatories are aligned at an angle of $45^\circ$.
Our findings could inform the design and deployment strategies of future GW observatories, enhancing their scientific output.

An aspect that is not discussed in this paper is the impact of environmental noise on the different configurations.
A specialty of the ET triangle configuration is the direct access to a GW signal-free null stream, which can be formed by a linear combination of the interferometer strain data~\cite{Regimbau:2012}.  While it does not directly improve inference~\cite{Goncharov:2022}, it provides information about the instrumental noise without contamination of GW signals and a way to extract the PSD of the detector noise~\cite{Wong:2021,Janssens:2022-2,Wu:2022}.
On the other hand, because the interferometers in the triangular design are located so close to each other, they are subject to correlated seismic and magnetic noise~\cite{Janssens:2021,Janssens:2022,Janssens:2024}. Since stochastic GWB searches rely on cross-correlating data, correlated noise could negatively impact the sensitivity to the stochastic background especially towards low frequencies. Moreover, Refs.~\cite{Cireddu:2023,Wong:2024} show that the effects of correlated noise should also be incorporated in CBC parameter estimation.

It would also be interesting to compare the capabilities of networks optimized for different targets with respect to more specific science cases, for instance the measurement of neutron star radii and probing the equation of state of dense nuclear matter, investigating physics near the black hole horizon or putting upper limits on different cosmological stochastic background scenarios.

\section*{Acknowledgments}

The authors thank Shubhanshu Tiwari and Alex Nitz for insightful comments and useful discussions. We thank John Veitch and Tito dal Canton for valuable discussions.

This work has been done thanks to the facilities offered by the Univ. Savoie Mont Blanc - CNRS/IN2P3 MUST computing center.

\bibliography{references}

\end{document}